\documentclass{ws-p9-75x6-50}
\def\beq{\begin{equation}}
\def\eeq{\end{equation}}
\def\be{\begin{eqnarray}}
\def\ee{\end{eqnarray}}
\def\ci{\cite}
\def\bi{\bibitem}

\def\vecq{{\bf q}}

\def\magq{|{\bf q}|}
\def\magp{|{\bf p}|}

\def\yt{{\widetilde y}}

\def\nut{{\widetilde \nu}}
\def\qt{{\widetilde q}}
\def\pt{{\widetilde p}}
\begin{document}

\title{Scaling in many-body systems and proton structure function}

\author{Omar Benhar}

\address{INFN, Sezione di Roma, I-00185 Roma, Italy\\
E-mail: benhar@roma1.infn.it}

\maketitle

\abstracts{
The observation of scaling in processes in which a
weakly interacting probe delivers large momentum ${\bf q}$ to a
many-body system simply reflects the dominance of incoherent scattering
off target constituents. While a suitably defined scaling function
may provide rich information on the internal dynamics of the target,
in general its extraction from the measured cross section requires careful
consideration of the nature of the interaction driving
the scattering process. The analysis of deep inelastic electron-proton
scattering in the target rest frame within standard many-body theory
naturally leads to the emergence of a scaling function that, unlike the
commonly used structure functions $F_1$ and $F_2$, can be directly identified
with the intrinsic proton response.          
}

\section{Introduction}

Scaling is observed in a variety of scattering
processes involving many-body systems \ci{west}. For example, at large 
momentum transfer $\magq$ the response of liquid helium measured 
by inclusive scattering of thermal neutrons, which in general depends upon 
{\it both} ${\bf q}$ and the energy transfer $\nu$, exhibits a striking scaling 
behavior, i.e. it becomes a function of the single variable 
$y = (m/\magq)(\nu - \vecq^2/2m)$, $m$ being the mass of the 
helium atom \cite{he}. Scaling in a similar 
variable occurs in inclusive electron-nucleus scattering 
at $\magq >$ 500 MeV and electron energy loss $\nu < Q^2/2M$, 
where $Q^2 = \magq^2  -\nu^2$ and $M$ is the nucleon mass \cite{nuclei}. 
Another most celebrated example is scaling of the deep inelastic proton 
structure functions, measured by lepton scattering at large $Q^2$,
in the Bjorken variable $x=Q^2/2M\nu$ \ci{bj}.

The relation between Bjorken scaling and scaling in the variable $y$, whose
definition and interpretation emerge in a most natural fashion from the
treatment of the scattering process within many-body theory in the 
target rest frame, has been discussed by many authors 
(see, e.g., ref.\ci{west89}).
Recently, deep inelastic data have been also shown to scale
in the variable $\yt = \nu - \magq$ \ci{bps}, related to
{\it both} $y$ {\it and} the Nachtman variable $\xi$ \ci{Nac}, which in 
turn coincides with $x$ in the $Q^2 \rightarrow \infty$ limit.

The fact that scaling is observed in processes driven by different 
interactions clearly indicates that its occurrence reflects 
the dominance of a common reaction mechanism, independent 
of the underlying dynamics. In all instances, scaling is 
indeed a consequence of the 
onset of the impulse approximation (IA) regime, in which scattering 
of a weakly interacting probe by a composite target reduces to the 
incoherent sum of elementary scattering processes involving its constituents. 

While the primary goal of scaling analysis is the identification of the 
dominant reaction mechanism, it has to be pointed out that the scaling variable 
has a straightforward physical interpretation, and a suitably defined 
scaling function, being directly related to the target response, 
contains a great deal of dynamical information. 

In general, extracting the target response from the mesasured 
cross section requires 
careful consideration of the nature of the interaction driving the scattering 
process. In neutron-liquid helium scattering, as the probe-constituent 
coupling is purely scalar, the cross section coincides with
the response up to a kinematical factor \ci{he}. On the other hand, 
to obtain the target response in the case of electron-nucleus scattering
 one has to divide out the elementary electron-nucleon cross section \ci{nuclei}.
In this paper we will discuss the application of the latter procedure to the 
analysis of 
deep inelastic scattering (DIS) of electrons by protons.
 
The theoretical treatment of scattering off a many-body system and the 
assumptions involved in the impulse approximation (IA) are described in 
Section 2, where the case of neutron-liquid helium scattering is 
considered as a pedagogical example. Section 3 is devoted to the application 
of the many-body approach to the more complex case of electron-proton scattering, 
as well as to the derivation of the appropriate scaling variable and scaling 
function. The relation between the analysis of DIS proposed in this paper
and the standard Bjorken scaling analysis is discussed in Section 4, 
where the implications of the differences between the two approaches are 
emphasized. Finally, Section 5 contains a summary and the conclusions.

\section{Scattering off many-body systems in the IA regime and
$y$-scaling}

Let us consider scattering off a norelativistic bound system consisting of N 
{\it pointilke scalar} particles of mass $m$, and assume that the 
probe-target interaction be weak, so 
that Born approximation can be safely used. The differential cross section
of the process in which a beam particle, carrying momentum ${\bf k}$ and
energy $E$, is scattered into the solid angle $d\Omega$ with energy 
$E^\prime = E - \nu$ and momentum ${\bf k}^\prime$ can then be written 
\beq
\frac{d\sigma}{d\Omega dE^\prime} = \frac{\sigma}{4 \pi}\ 
\frac {|{\bf k}^\prime|}{|{\bf k}|}
S({\bf q},\nu)\ ,
\label{xsec:1}
\eeq
where $\sigma$ is the probe-constituent total cross section and the 
{\it response function} $S({\bf q},\nu)$, containing all the information 
on the structure of the target, is defined as
\beq
S({\bf q},\nu) = \sum_n | \langle n |\rho_{\bf q}| 0 \rangle |^2
\delta(\nu+E_0+E_n) 
 = \int \frac{dt}{2\pi}\ {\rm e}^{i\nu t}\
\langle 0 | \rho^\dagger_{\bf q}(t) \rho_{\bf q}(0) | 0 \rangle\ .
\label{response:12}
\eeq
In the above equations, $| 0 \rangle$ and $| n \rangle$ are the target
ground and final state, satisfying 
$H | 0 \rangle=E_0| 0 \rangle$ and $H | n \rangle=E_n| n \rangle$, 
$H$ being the hamiltonian describing the internal target dynamics,
 $\rho_{\bf q}(t) = {\rm e}^{iHt}\rho_{\bf q}{\rm e}^{-iHt}$, with  
 $\rho_{\bf q} = \sum_{\bf k} a^\dagger_{{\bf k}+{\bf q}}a_{\bf k}$,
 and $a^\dagger_{\bf k}$ and $a_{\bf k}$ are constituent creation and
annihilation operators, respectively.

Rewriting Eq.(\ref{response:12}) in coordinate space leads to
the expression
\beq
S({\bf q},\nu)=\sum_n \left| \int dR\  \langle n | R \rangle 
\sum_{i=1}^{\rm N} 
{\rm e}^{i {\bf q}\cdot{\bf r}_i}\  \langle R | 0 \rangle \right|^2
\delta(\nu+E_0+E_n)\ ,
\label{resp:coord}
\eeq
where $R\equiv({\bf r}_1,\ldots,{\bf r}_{\rm N})$ specifies the target 
configuration, while $\langle R | 0 \rangle$ and
$\langle R | n \rangle$ denote its initial and final state wave 
functions.
 
The main assumption underlying IA is that, as
the space resolution of a probe delivering momentum $\vecq$ 
is $\sim 1/\magq$, at large enough $\magq$ (typically $\magq \gg 2\pi/d$, 
$d$ being the average separation between target constituents) the 
target is seen 
by the probe as a collection of individual particles.
In addition, final state
interactions (FSI) between the hit constituent, carrying a large momentum 
$\sim {\bf q}$, and the residual 
(N-1)-particle system are assumed to be negligibly small.

In the IA regime the scattering process reduces to the 
incoherent sum 
of elementary processes involving only one constituent, the remainig 
(N-1) particles acting as spectators, and Eq.(\ref{resp:coord}) simplifies to 
\beq 
S({\bf q},\nu)= \sum_i \sum_n \left| \int dR\ \langle n | R \rangle\  
{\rm e}^{i {\bf q}\cdot{\bf r}_i}\  \langle R | 0 \rangle \right|^2
\delta(\nu+E_0+E_n)\ .
\label{resp:incoh}
\eeq
The IA final state $|n\rangle$, carrying total momentum ${\bf q}$, has the 
structure
\beq
| n \rangle = | i,{\cal R} \rangle = |i({\bf p}^\prime) \rangle \otimes 
| {\cal R}({\bf q} - {\bf p}^\prime) \rangle\ ,
\label{state:IA}
\eeq
its energy being given by $E_n = E_{{\bf p}^\prime} + E_{{\cal R}}$,
where $E_{{\bf p}^\prime}=|{\bf p}^\prime|^2/2m$ and 
$E_{{\cal R}}$ denote the energies of the 
free struck 
constituent, with momentum ${\bf p}^\prime$, and the spectator 
(N-1)-particle system, with momentum ${\bf q} - {\bf p}^\prime$, 
respectively. As a consequence, the sum over final states can 
be carried out replacing
\beq
\sum_n | n \rangle \langle n | \rightarrow \int d^3 p^\prime\ 
| i({\bf p}^\prime) \rangle \langle i({\bf p}^\prime) |\ \sum_{\cal R}\ 
| {\cal R}({\bf q} - {\bf p}^\prime) \rangle 
\langle {\cal R}({\bf q} - {\bf p}^\prime) |\ . 
\label{sum:IA}
\eeq
Substitution of Eqs.(\ref{state:IA}) and (\ref{sum:IA}) 
into Eq.(\ref{resp:incoh}) yields (see, e.g., ref.\ci{impulse})
\beq
S({\bf q},\nu) = \sum_i \int d^3 p \int d p_0\ 
P_i({\bf p},p_0) \delta(\nu + p_0 - E_{{\bf p}+{\bf q}})\ ,
\label{resp:IA}
\eeq
where the function
\beq 
P_i({\bf p},p_0) = \sum_{\cal R} \left| 
\langle 0 | i,{\cal R} \rangle \right|^2 \delta(p_0 + E_{\cal R} - E_0)
\label{spec:fcn}
\eeq 
gives the probability of finding the $i$-th constituent with momentum 
${\bf p}$ and energy $p_0$ in the target ground state.

Eqs.(\ref{resp:IA}) and (\ref{spec:fcn}) show that the IA response 
only depends upon ${\bf q}$ and $\nu$ through the energy-conserving 
$\delta$-function, requiring 
\beq
\nu + E_0 - E_{\cal R} + E_{{\bf p}+{\bf q}} = 0\ .
\label{energy:cons}
\eeq
The occurrence of scaling, i.e. the fact that, up to a kinematical factor
$K(\magq,\nu)$, $S({\bf q},\nu)$ becomes a function of a single variable, 
simply reflects the 
fact that in the IA regime, in which energy conservation is expressed by 
Eq.(\ref{energy:cons}), ${\bf q}$ and $\nu$ are no longer 
independent variables. One can then define a new variable
$y=y({\bf q},\nu)$ such that, as $\magq \rightarrow \infty$, 
 $K(\magq,\nu) S({\bf q},\nu) \rightarrow F(y)$.

In the simple case of neutron scattering off liquid helim, in which 
the cross section can be written exactly as in Eq.(\ref{xsec:1}) 
and the energy dependence of $P_i({\bf p},p_0)$ can be safely neglected
\cite{impulse}, Eq.(\ref{energy:cons}) takes the form 
\beq
\nu + \frac{{\bf p}^2}{2 m} - \frac{|{\bf p}+{\bf q}|^2}{2 m} = 0\ ,
\eeq
$m$ being the mass of the helium atom. It follows that, defining 
\beq
y = \frac{m}{\magq} \left( \nu - \frac{\magq^2}{2m} \right)\ ,
\label{he:y}
\eeq
in the $\magq \rightarrow \infty$ limit  
\beq
\frac{\magq}{m}\  S({\bf q},\nu) \rightarrow F(y)\ .
\label{he:fy}
\eeq

\begin{figure}[ht]
\begin{center}
\epsfxsize=20.0pc
\epsfbox{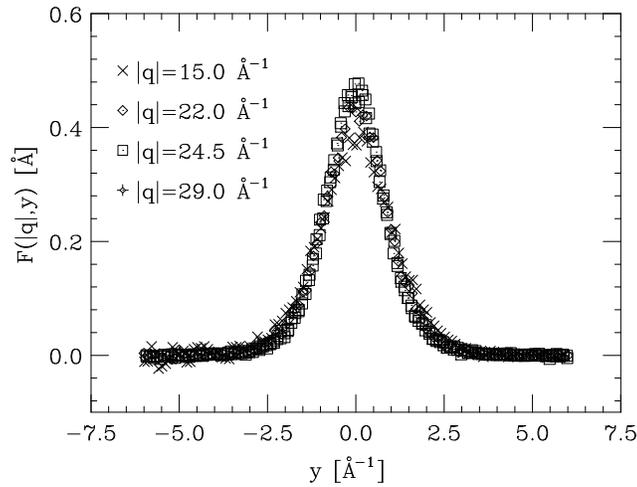}
\caption{
Scaling functions $F(y)$, defined as in Eq.(\protect\ref{he:fy}), measured
by neutron scattering off superfluid $^4$He at T$=$1.6\
$^\circ$K \protect\ci{scal:he}.
}
\label{helium}
\end{center}
\end{figure}

Fig.\ref{helium} shows the behavior of $F(y)$, defined by the 
above equation, measured in neutron scattering off
superfluid $^4$He at T $=$1.6 $^\circ$K \cite{scal:he}.  
The curves corresponding to $F(y)$ at $\magq  \geq$ 15 \AA$^{-1}$ lie 
on top of one another, clearly 
indicating the onset of the scaling regime.

The variable $y$ defined by Eq.(\ref{he:y}) does have a straightforward
physical interpretation, and so does the scaling function of Eq.(\ref{he:fy}).
 The scaling variable can be identified with the initial longitudinal momentum 
of the struck atom, $p_\parallel$, while $F(y)$ can be related to the momentum 
distribution, $n(\magp)$. In fact, the connection linking the response in 
the scaling regime to $n(\magp)$ has been been 
extensively exploited to obtain momentum distributions of normal and 
superfluid $^4$He from neutron scattering data \ci{he2}. 

\section{Scaling in deep inelastic scattering and proton response} 

The unpolarized electron-proron ($ep$) scattering cross section 
is usually written in the form (see, e.g., ref.\ci{close})
\beq
\frac{d^2 \sigma}{d\Omega dE^\prime} = \frac{\alpha^2}{Q^4}\
\frac{E^\prime}{E}\ L_{\mu \nu} W^{\mu \nu}\ ,
\label{sigmaep:1}
\eeq                            
where $E$ and $E^\prime$ denote the initial and final electron energy, 
respectively, $Q^2=-q^2$, $q = k-k^\prime \equiv (\nu,\vecq)$ being the 
four momentum transfer, $\alpha$ is the fine structure constant and 
$L_{\mu \nu}(k,k^\prime)$ is the
tensor associated with the electron, fully specified by the measured 
kinematical variables. 

The information on the structure of the target is contained 
in the tensor $W_{\mu \nu}$. It can be written in a form reminiscent 
of Eq.(\ref{response:12}), in which the density 
fluctuation operator $\rho_{\bf q}$ is replaced by the proton electromagnetic 
current $J_\mu$ and the relativistic nature of the process is properly taken
into account:
\beq
W_{\mu \nu} = \sum_n 
\langle 0 | J_\mu^\dagger | n \rangle \langle n | J_\nu | 0 \rangle
  \delta^{(4)}(P_0+q-P_n)
= \int \frac{d^4x}{(2\pi)^4}\ {\rm e}^{iqx}\  
\langle 0 | J_\mu^\dagger(x) J_\nu(0) | 0 \rangle\ ,
\label{Wmunu:12}
\eeq
where $P_0 \equiv (M,{\bf 0})$, $M$ being the proton mass, and 
$P_n$ are the four-momenta of the initial and final hadronic states.

We will now make the assumption that the proton can be viewed as 
a many-body system, consisting of {\it bound} pointlike Dirac particles 
of mass $m$, 
and proceed to apply the analysis described in Section 2 to $ep$ 
scattering. This will of course amount to assume, as in 
the standard parton model of DIS \ci{close},
that over the short spacetime scale relevant to the scattering process confinement
does not play a role, so that proton constituents can be described in terms of
physical states. It has to be emphasized, however, that, unlike the parton 
model, the present approach {\it does not} involve the
additional requirement that proton constituents be on mass shell. 

In the IA scheme the target tensor given by Eq.(\ref{Wmunu:12}) 
 is replaced by a weighted sum of tensors 
describing the electromagnetic structure of proton constituents
(compare to Eq.(\ref{resp:IA})):
\beq
W^{\mu\nu} \rightarrow  \sum_i q_i^2\  
\int d^4 p\ P_i(p)\ \frac{1}{2E_{\bf p}}\ 
\frac{1}{2E_{{\bf p}+{\bf q}}}
w^{\mu\nu}(p,q)\  \delta \left( \nu + p_0 - E_{{\bf p}+{\bf q}} \right)\ ,
\label{tensor:IA}
\eeq
where $E_{\bf p} = \sqrt{|{\bf p}|^2 + m^2}$, 
$E_{{\bf p}+{\bf q}} = \sqrt{|{\bf p}+{\bf q}|^2 + m^2}$ and
$P_i(p)$ is the probability that the $i$-th constituent,
 whose charge in units of the electron charge is denoted $q_i$, carry
four-momentum $p\equiv(p_0,{\bf p})$.  As a consequence, the cross section of 
Eq.(\ref{sigmaep:1}) can be cast in the form
\beq 
\frac{d\sigma}{d\Omega dE^\prime} = \sum_i q_i^2\  \int d^4 p\ P_i(p)\ 
\left( \frac{d\sigma_c}{d\Omega dE^\prime} \right)
 \delta \left( \nu + p_0 - E_{{\bf p}+{\bf q}} \right)\ ,
\label{sigmaep:IA}
\eeq
$(d\sigma_c/d\Omega dE^\prime)$ being the elementary 
electron-constituent cross section. 

Substitution of Eq.(\ref{tensor:IA}) into Eq.(\ref{sigmaep:1}) and 
comparison with Eq.(\ref{sigmaep:IA}) leads to
\beq
\left( \frac{d\sigma_c}{d\Omega dE^\prime} \right) = 
\frac{\alpha^2}{Q^4}\ \frac{E^\prime}{E}\ 
\frac{1}{2E_{\bf p}}\ \frac{1}{2E_{{\bf p}+{\bf q}}}
L_{\mu \nu} w^{\mu \nu}\ ,
\label{sigmac:1}
\eeq
where the tensor $w^{\mu \nu}$ is defined as
\beq
w^{\mu \nu} = 2\left\{ \pt^\mu (\pt + \qt)^\nu 
+ \pt^\nu (\pt + \qt)^\mu - g^{\mu \nu} \left[ \left( \pt (\pt +\qt) \right) 
- m^2 \right] \right\}\ .
\label{wmunu}
\eeq
with $\pt\equiv(E_{\bf p},{\bf p})$, 
$\qt\equiv(\nut,{\bf q})$ and $\nut = E_{{\bf p}+{\bf q}} - 
E_{\bf p} = \nu + p_0 - E_{\bf p}$. 
The above equations show that the formalism of IA allows one to describe
scattering off a {\it bound} constituent in terms of the electromagnetic 
tensor associated with a {\it free} Dirac particle of inital four-momentum 
${\widetilde p}$ and final four-momentum 
$p^\prime = {\widetilde p} + \qt = p + q$. It has to be mentioned,  
however, that, on account of the replacement $q \rightarrow \qt$, 
$w^{\mu \nu}$ defined by Eq.(\ref{wmunu}) is
manifestly {\it non} gauge-invariant, i.e. $q_\mu w^{\mu \nu} \neq 0$. 
Gauge invariance can be restored 
using a somewhat {\it ad hoc} prescription, originally proposed 
by De Forest \ci{defo} and widely 
used in the theoretical analysis of electron-nucleus scattering processes. 
Although violation of gauge invariance is in general an unpleasant feature
inherent in the IA scheme, it does not play a quantitatively significant role 
in the context of DIS, as the non 
gauge-invariant contributions to the elementary cross section become 
vanishingly small in the $\magq \rightarrow \infty$ limit.

Using Eqs.(\ref{sigmac:1}) and (\ref{wmunu}) and following ref.\ci{defo} one 
gets
\beq
\left( \frac{d\sigma_c}{d\Omega dE^\prime} \right) = 
\left( \frac{d\sigma}{d\Omega} \right)_M 
\left[ \sigma_2 + 2 \sigma_1\ \tan^2 \frac{\theta}{2} \right]\ ,
\label{sigdefo}
\eeq
where  $(d\sigma/d\Omega)_M$ denotes the Mott cross section, $\theta$ is 
the electron scattering angle, 
\beq
\sigma_1 = \frac{1}{ E_{\bf p} E_{{\bf p}+{\bf q}} }\ 
\left( - \frac{\qt^2}{4} + \frac{p_\perp^2}{2} \right)\ ,
\label{sig:1}
\eeq
and
\beq
\sigma_2 = \frac{1}{ E_{\bf p} E_{{\bf p}+{\bf q}} }
\left( \frac{q^2}{\magq^2} \right)\ \left[
- \frac{\qt^2}{4} \left( \frac{q^2}{\qt^2} - 1 \right)
+ \left( \frac{q^2}{\magq^2} \right) 
\left( \frac{ E_{\bf p} + E_{{\bf p}+{\bf q}} }{2} \right)^2 
-\frac{p_\perp^2}{2} \right]\ ,
\label{sig:2}
\eeq
$p_\perp$ being the component of the constituent momentum perpendicular to 
the momentum transfer.

Substitution Eqs.(\ref{sigdefo})-(\ref{sig:2}) into Eq.(\ref{sigmaep:1})
leads to the familiar expression of the $ep$ cross section in terms of 
the two structure functions $W_1$ and $W_2$: 
\beq
\frac{d^2\sigma}{d\Omega dE^\prime} =  
\left( \frac{d\sigma}{d\Omega} \right)_M \left[ W_2(|{\bf q}|,\nu) +
2 W_1 (|{\bf q}|,\nu) \tan^2 \frac{\theta}{2}  \right]\ ,
\label{sigmaep:2}
\eeq
with 
\beq
W_{1,2} = \sum_i q_i^2\  \int d^4 p\ P_i(p)\
\sigma_{1,2}(p,q)\ 
\delta \left( \nu + p_0 - \sqrt{ |{\bf p}+{\bf q}|^2 + m^2} \right)\ .
\label{W12:IA}
\eeq

Comparison between Eq.(\ref{sigmaep:1}) and Eq.(\ref{xsec:1}) 
shows that, due to the complexity of the electromagnetic
interaction, in the case of electron scattering the quantity carrying 
the information on the structure of the target, i.e. its intrinsic 
response, no longer 
appears as a multiplicative factor in the cross section. 
The electron scattering cross section is indeed 
{\it not} trivially related to the target response. 
For example, while the response of a many-body system to a probe 
delivering momentum ${\bf q}$ is in general
nonzero in the region $\nu \geq \magq$ \ci{bps,mark}, inaccessible to
electron scattering, at $\nu = \magq$ the structure function $W_2$ 
of Eq.(\ref{sigmaep:2}) vanishes, as required  by gauge
invariance, while $W_1$, proportional to the photoabsorption 
cross section, does not contribute to the cross section. 

The above problem can be circumvented if, as in the case of 
electron-nucleus scattering \ci{nuclei}, the momentum 
and energy dependence of the electron-constituent 
cross section is much weaker than that exhibited by the distribution 
function $P_i(p)$, which is a rapidly decreasing function of 
$\magp$ and the (positive) constituent binding energy $B=-p_0$. Under 
this assumption, the elementary cross section, 
evaluated at a constant $p={\overline p}$ corresponding to the 
maximum of $P_i(p)$, can be moved out of the integral 
in Eq.(\ref{sigmaep:IA}) and the proton response can be obtained from
\beq
S({\bf q},\nu) = \frac{d\sigma}{d\Omega dE^\prime} \left/
\left( \frac{d\sigma_c}{d\Omega dE^\prime}  \right)_{p={\overline p}}\ .
\right.
\label{proton:resp}
\eeq

Having identified the target response, we can now proceed to define the 
scaling variable exploiting energy conservation.
The requirement 
\beq
\nu + p_0 - \sqrt{|{\bf p}+{\bf q}|^2 + m^2} =
\nu + p_0 - p_\parallel - \magq + {\cal O}\left(\frac{1}{\magq}\right) = 0\ ,
\label{en:cons}
\eeq
where $p_0 = M - E_{\cal R}$ and $E_{\cal R} = \sqrt{\magp^2 + M_{\cal R}^2}$, 
$M_{\cal R}$ being the mass of the spectator system, implies that, as 
$\magq \rightarrow \infty$, the quantity
\beq
\yt = \nu - \magq = p_\parallel + p_0
\label{def:yt}
\eeq
becomes independent of ${\bf q}$. Hence, in this limit 
$S({\bf q},\nu)$ exhibits scaling in the variable $\yt$, i.e.
\beq
S({\bf q},\nu) \rightarrow F(\yt). 
\label{F:DIS}
\eeq

It has to be pointed out that $\yt$ {\it does not} have the same physical 
interpretation as the variable $y$ defined in Section 2, as it {\it does not} 
coincide with the constituent longitudinal momentum. Obviously, as 
$p_0$ is independent of $q$, scaling in $y$ necessarily implies scaling 
in $\yt$, and {\it viceversa}. The choice of $\yt$ as scaling 
variable in DIS is motivated by the fact that 
$\yt$ is trivially related to another variable commonly used 
in the same context, the Nachtman variable $\xi$ \ci{Nac},
through
\beq
- \frac{\yt}{M} = \frac{2x}{1+\sqrt{1+4M^2x^2/Q^2}} = \xi \ ,
\label{yt:csi}
\eeq
where $x$ is the Bjorken variable. The above equation shows that 
in the $Q^2 \rightarrow \infty$ limit $\yt$ coincides with $x$. It has to 
be noticed that the IA scheme provides a simple 
physical intepretation of Nachtman's variable, whose definition was
originally obtained in a totally different fashion. 

\section{$\yt$-scaling analysys of DIS data}

According to the IA picture, the $\yt$-scaling function can be obtained dividing 
either structure function by the appropriate contribution to the elementary
cross section. In fact, from Eqs.(\ref{sigmaep:2})-(\ref{proton:resp}) and 
(\ref{F:DIS}) it follows that, in the $\magq \rightarrow \infty$ limit, 
\beq
S({\bf q},\nu) = \frac{W_1}{{\overline \sigma}_1} 
= \frac{W_2}{{\overline \sigma}_2} \rightarrow F(\yt)  \ ,
\eeq
where ${\overline \sigma}_{1,2}=(\sigma_{1,2})_{p={\overline p}}$, 
with ${\overline p}\equiv(B_0,{\bf p}_{min})$, $B_0$ is
the minimum constituent binding energy and the magnitude of the 
minimum momentum allowed
in the kinematics specified by ${\bf q}$ and $\nu$ is 
\beq
|{\bf p}_{min}| = \frac{1}{2} \left| 
\frac{M_{\cal R}^2 - (\yt + M)^2}{\yt + M} \right|\ ,
\label{p:min}
\eeq
with $M_{\cal R} = M - m + B_0$.

\begin{figure}[ht]
\begin{center}
\epsfxsize=20.0pc
\epsfbox{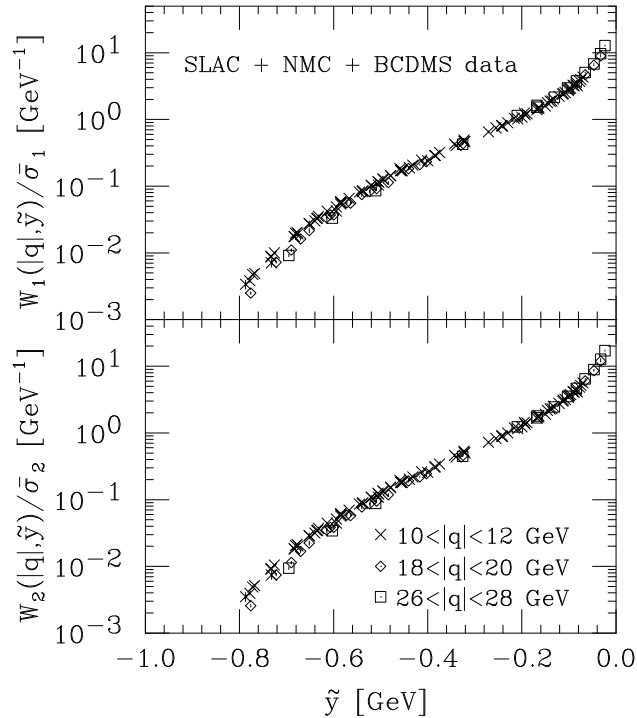}
\caption{
Scaling functions obtained from $W_1/{\overline \sigma}_1$ (upper panel)
and $W_2/{\overline \sigma}_2$ (lower panel). The experimental
structure functions are from refs.\protect\ci{SLAC,NMC,BCDMS}, whereas
the elementary cross sections have been evaluated
using Eqs.(\protect\ref{sig:1}) and (\protect\ref{sig:2}), with $m = 300$
MeV and $B_0 = 200$ MeV.
}
\label{DIS:1}
\end{center}
\end{figure}

The scaling analysis in terms of $\yt$ involves two 
parameters: the constituent mass, $m$, and binding energy, $B_0$, or 
equivalently $m$ and the mass of the spectator system, $M_{\cal R}$. 
Fig. \ref{DIS:1} shows the quantities $W_1/{\overline \sigma}_1$ 
(upper panel) and $W_2/{\overline \sigma}_2$ (lower panel),  
obtained from data taken at SLAC \ci{SLAC} and CERN \ci{NMC,BCDMS} and
rearranged in bins of constant $\magq$ centered at 11, 19 and 27 GeV, 
plotted as a function of $\yt$. The structure functions $W_1$ have been
obtained using the parametrization of $R=\sigma_L/\sigma_T$ of 
ref. \ci{RLT}, while the elementary cross sections have been evaluated
from Eqs.(\ref{sig:1}) and (\ref{sig:2}), with $m = 300$ MeV and 
$B_0 = 200$ MeV. It clearly appears that in both cases scaling sets in at 
$\magq >$ 10 GeV.

\begin{figure}[ht]
\begin{center}
\epsfxsize=20.0pc
\epsfbox{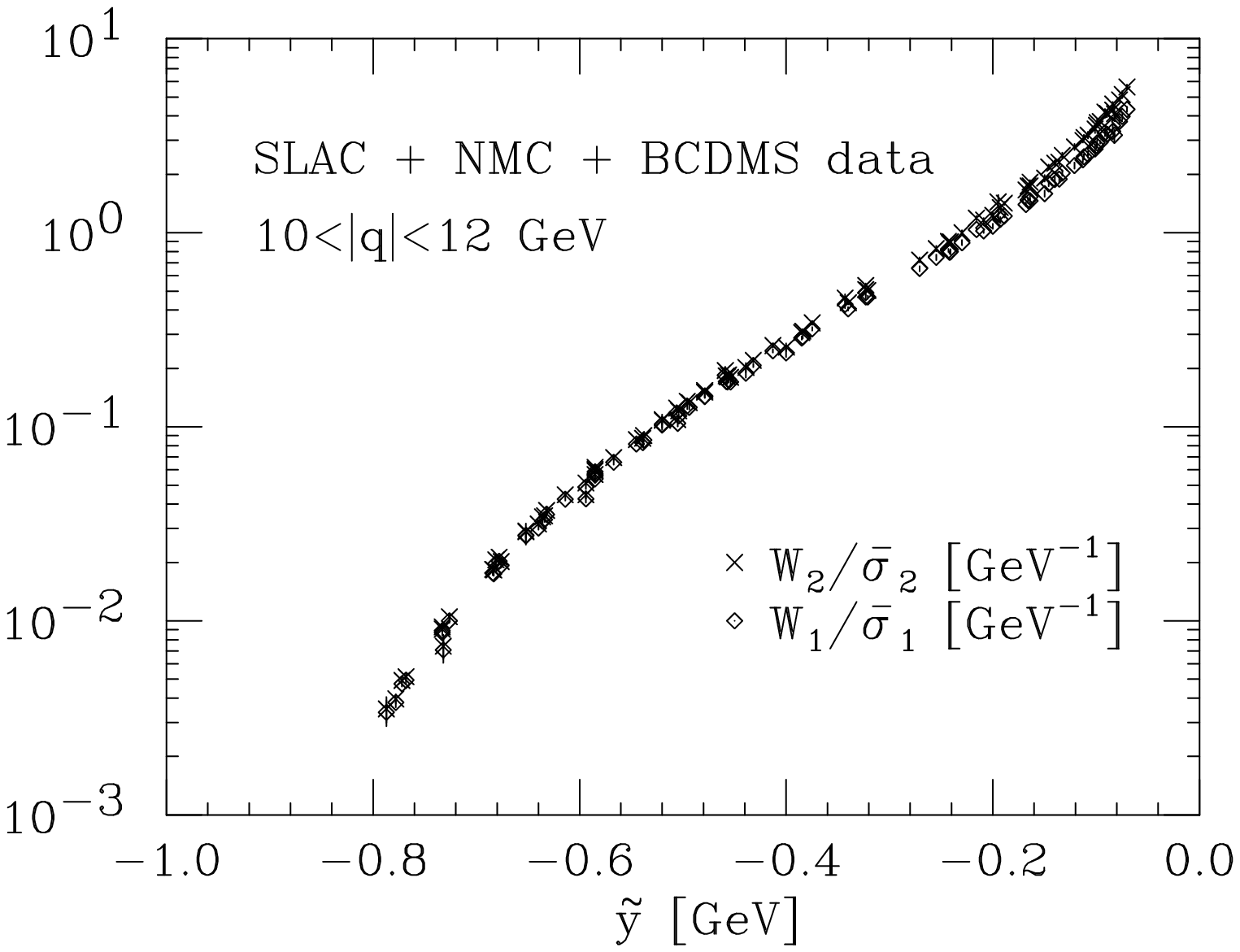}
\caption{
Comparison between $W_1/{\overline \sigma}_1$ (diamonds) and
$W_2/{\overline \sigma}_2$ (crosses) of fig. \protect\ref{DIS:1}
at 10 $\leq \magq \leq$ 12 GeV.
}
\label{DIS:2}
\end{center}
\end{figure}

The second feature predicted by the IA analysis, i.e. that 
$W_1/{\overline \sigma}_1$ and $W_2/{\overline \sigma}_2$ scale to 
the {\it same} function $F(\yt)$ is illustrated in fig. \ref{DIS:2}.

\begin{figure}[ht]
\begin{center}
\epsfxsize=20.0pc
\epsfbox{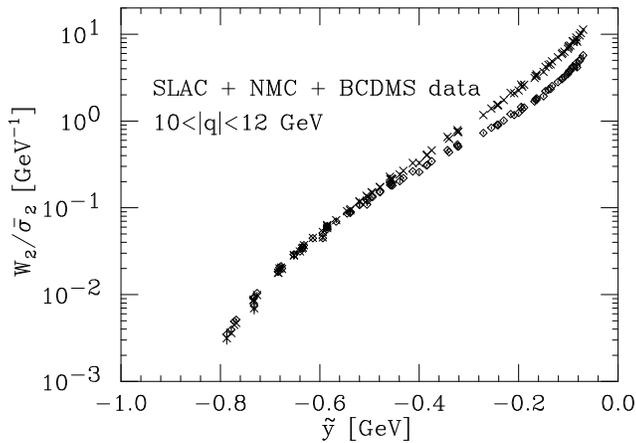}
\caption{
Dependence of the $\yt$-scaling function upon the mass of the
spectator system $M_{\cal R}$. Crosses and diamonds correspond
to $M_{\cal R}=$ 638 and 838 MeV, respectively. The constituent mass
is $m$ = 300 MeV.
}
\label{DIS:3}
\end{center}
\end{figure}

It has to be pointed out that the occurrence of $\yt$ scaling {\it does not} 
depend upon the values of either $m$ or $B_0$. In particular, it does not 
require that the constituent mass be vanishingly small, nor that the 
constituent be on mass shell. Varying $m$ and $B_0$ in the range 
10-300 MeV does not affect the scaling behavior displayed in figs. \ref{DIS:1} 
and \ref{DIS:2}. However, it does affect the scaling function 
extracted from the data. While in general different combinations of $m$ and
$B_0$ correspond to different $F(\yt)$'s, the scaling function turns out 
to be only sensitive to the difference $m-B_0$, i.e. to the mass of the 
spectator system $M_{\cal R}$. Fig. \ref{DIS:3} shows that increasing 
$M_{\cal R}$ results in a shift of $F(\yt)$ towards larger values of $\yt$
 at $\yt > -0.5$ GeV. 

DIS data are usually analyzed in terms of the two dimensionless 
structure functions $F_1 = M W_1$ and $F_2 = \nu W_2$. In the Bjorken 
limit $Q^2, \nu \rightarrow \infty$, with $Q^2/\nu$ finite 
and $\nu/\magq \rightarrow 1$, both $F_1$ 
and $F_2$ exhibit scaling in the variable $x$, whose definition 
in the target rest frame is $x=Q^2/2M\nu$. 

The discussion of the previous Sections shows that, while $\yt$ coincides
with $x$ in the Bjorken limit (see eq.(\ref{yt:csi})), the $\yt$-scaling function 
cannot be identified with either $F_1$ or $F_2$. In fact, the structure of 
$F(\yt)$ is dictated by 
target dynamics only, whereas both $F_1$ and $F_2$ depend upon 
the elecron-constituent coupling as well. In general the scaling behavior 
displayed by 
$F_1$ and $F_2$ is a consequence of the fact that, in addition to the proton 
response, the quantities ${\overline \sigma}_1$ and 
$\nu {\overline \sigma}_2$ also scale.

$x$- and $\yt$-scaling analyses can be reconciled making the standard 
assumption of parton model that the constituent mass be vanishingly small.
In fact, in the $m \rightarrow 0$ limit $\sigma_1 \equiv 1$ and 
$\sigma_2 = Q^2/\magq^2$, implying in turn $F_1 = MF(\yt)$ and 
$F_2 =  \nu Q^2 F(\yt)/\magq^2 = 2xF_1$. However, it has to be 
emphasized that, in spite of
the fact that in textbook derivations the requirement $m \sim 0$ is 
often introduced as a necessary condition for Bjorken scaling 
(see, e.g., ref. \ci{close}), the present analysis shows that scaling occurs 
irrespective of the constituent mass.

\section{Conclusions}

The results described in this paper show that the IA analysis of scattering 
processes off many-body targets, successfully employed to describe 
neutron-liquid helium and electron-nucleus data, can be extended to the 
case of DIS of electrons by protons.

While the scaling variable of many-body theory, whose definition and 
intepretation naturally emerge in the target rest frame, is trivially 
related to the Bjorken variable $x$, $\yt$-scaling analysis differs
from the standard parton model of DIS in that it does not make any 
assumptions on the mass and binding energy of proton constituents. The
occurrence of $\yt$-scaling turns out to be a mere consequence of the 
onset of the IA regime, and cannot be taken as unambiguous evidence of 
scattering off massless, on shell constituents.

The $\yt$-scaling function also differs from the commonly used structure 
functions $F_1$ and $F_2$, and can be directly 
identified with the intrinsic proton response. The dependence of $F(\yt)$ upon the constituent binding energy 
implies that, if the interactions between proton constituents are strong 
enough to shift a fraction of the strength into the timelike region 
$\nu > \magq$, the proton response extracted from electron scattering data 
does not fulfill the constituent number sum rule \cite{sumrule}. The occurrence
of strength located in the region unaccessible to electron scattering is 
a well known feature of the response of interacting many-body systems. One 
that has long been recognized as the reason why the integral of the 
charge response measured by electron-nucleus scattering does not yield
the total nuclear charge Z \ci{book}. Recently, it has been also shown that
about 10\% of the response of a scalar relativistic particle 
bound in a linear confining 
potential resides in the timelike region \ci{mark}.

As a final remark, it has to be mentioned that the approach discussed in this
paper, showing that scaling is by no means incompatible with massive 
constituents, suggests that the ability of the constituent quark model 
of the proton to describe DIS should be reconsidered, carrying out a 
$\yt$-scaling analysis in which finite mass effects are consistently taken 
into account in both the ptoton response and the electron-constituent
interaction vertex. 

\section*{Acknowledgments}
The author is deeply indebted to Vijay R. Pandharipande and Ingo Sick for
many illuminating discussions on different issues related to the
subject of this paper.    


\end{document}